\documentclass[aps,twocolumn,prl,superscriptaddress,preprintnumbers,amsmath,showpacs,tightenlines]{revtex4}
\usepackage{graphicx}
\usepackage{dcolumn}
\usepackage{bm}
\usepackage{amssymb}
\usepackage{amssymb}
\usepackage{epsfig,graphicx,times}
\def\>{\rangle}

\begin{document}

\title{Chaos can act as a decoherence suppressor}

\author{Jing Zhang}\email{jing-zhang@mail.tsinghua.edu.cn}
\affiliation{Department of Automation, Tsinghua University,
Beijing 100084, P. R. China} \affiliation{Center for Quantum
Information Science and Technology, Tsinghua National Laboratory
for Information Science and Technology, Beijing 100084, P. R.
China} \affiliation{Department of Physics and National Center for
Theoretical Sciences, National Cheng Kung University, Tainan
70101, Taiwan}
\author{Yu-xi Liu}
\affiliation{Institute of Microelectronics, Tsinghua University,
Beijing 100084, P. R. China} \affiliation{Center for Quantum
Information Science and Technology, Tsinghua National Laboratory
for Information Science and Technology, Beijing 100084, P. R.
China}
\author{Wei-Min Zhang}
\affiliation{Department of Physics and Center for Quantum
Information Science, National Cheng Kung University, Tainan 70101,
Taiwan}
\author{Lian-Ao Wu}
\affiliation{Department of Theoretical Physics and History of
Science, The Basque Country University (EHU/UPV) and IKERBASQUE -
Basque Foundation for Science, 48011, Bilbao, Spain}
\author{Re-Bing Wu}
\affiliation{Department of Automation, Tsinghua University,
Beijing 100084, P. R. China} \affiliation{Center for Quantum
Information Science and Technology, Tsinghua National Laboratory
for Information Science and Technology, Beijing 100084, P. R.
China}
\author{Tzyh-Jong Tarn}
\affiliation{Department of Electrical and Systems Engineering,
Washington University, St.~Louis, MO 63130, USA}
\affiliation{Center for Quantum Information Science and
Technology, Tsinghua National Laboratory for Information Science
and Technology, Beijing 100084, P. R. China}
\affiliation{Department of Physics and National Center for
Theoretical Sciences, National Cheng Kung University, Tainan
70101, Taiwan}

\date{\today}

\begin{abstract}
We propose a strategy to suppress decoherence of a solid-state
qubit coupled to non-Markovian noises by attaching the qubit to a
chaotic setup with the broad power distribution in particular in
the high-frequency domain. Different from the existing decoherence
control methods such as the usual dynamics decoupling control,
high-frequency components of our control are generated by the
chaotic setup driven by a low-frequency field, and the generation
of complex optimized control pulses is not necessary. We apply the
scheme to superconducting quantum circuits and find that various
noises in a wide frequency domain, including low-frequency $1/f$,
high-frequency Ohmic, sub-Ohmic, and super-Ohmic noises, can be
efficiently suppressed by coupling the qubits to a Duffing
oscillator as the chaotic setup. Significantly, the decoherence
time of the qubit is prolonged approximately $100$
times in magnitude.
\end{abstract}

\pacs{05.45.Gg, 03.67.Pp, 05.45.Mt}

\maketitle

\emph{Introduction.---} Solid state quantum information
processing~\cite{You} develops very rapidly in recent years. One
of the basic features that makes quantum information unique is the
quantum parallelism resulted from quantum coherence and
entanglement. However, the inevitable interaction between the
qubit and its environment leads to qubit-environment entanglement
that deteriorates quantum coherence of the qubit. In solid state
systems, the decoherence process is mainly caused by the
non-Markovian noises induced, e.g., by the two-level fluctuators
in the substrate and the charge and flux noises in the
circuits~\cite{Tsai,Shiokawa,Bergli}.

There have been numbers of proposals for suppressing non-Markovian
noises in solid-state systems. Most of them suppress noises in a
narrow frequency domain, e.g., low-frequency
noises~\cite{Tsai,Shiokawa}. Among the proposed decoherence
suppression strategies, the dynamical decoupling control (DDC)
~\cite{Viola1} is relatively successful in suppressing
non-Markovian noises in a broad frequency domain and has recently
been demonstrated in the solid-state system
experimentally~\cite{JFDu}. The main idea of the DDC is to utilize
high frequency pulses to flip states of the qubit rapidly,
averaging out the qubit-environment coupling. The higher the
frequency of the control pulse is, the better the decoherence
suppression effects are. Efforts have been made to optimize the
control pulses~\cite{LAWu} in the DDC, however, the requirements
of generating extremely high frequency control pulses or complex
optimized pulses make it difficult to be realized in solid-state
quantum information system, in particular in superconducting
circuits.

In this letter, we propose a method to extend the decoherence time
of the qubit by coupling it to a chaotic setup~\cite{Pozzo}.
Although it is widely believed that the chaotic dynamics induces
inherent decoherence~\cite{WGWang,Peres,Jalabert,Vanicek,Jacquod},
e.g., the quantum Loschmidt echo~\cite{Jalabert}, we find
surprisingly that the frequency shift of the qubit induced by the
chaotic setup, which has not drawn enough attention in the
literature, can help to suppress decoherence of the qubit.

The main merits of this method are: (1) the high frequency components,
which contribute to the suppression of the non-Markovian noises,
can be generated by the chaotic setup even driven by a
low-frequency field; and (2) generating complex optimal control
pulses is not necessary.

\emph{Decoherence suppression by chaotic signals.---}  Consider
the following system-environment model~\cite{Bergli}
\begin{equation}\label{Hamiltonian of the total system}
\hat{H}=\left(\omega_q+\delta_q\right)\hat{S}_z+\sum_i\omega_i\hat{\sigma}_{z}^{(i)}+\sum_i\left(g_i\hat{\sigma}_{+}^{(i)}\hat{S}_-+{\rm
h.c.}\right),
\end{equation}
where $\omega_q$ ($\omega_i$), $\hat{S}_z$
($\hat{\sigma}_{z}^{(i)}$), and $\hat{S}_{\pm}$
($\hat{\sigma}_{\pm}^{(i)}$) are the angular frequency, the
$z$-axis Pauli operator, and the ladder operators of the qubit
(the $i$-th two-level fluctuator in the environment); $g_i$ is the
coupling constant between the qubit and the $i$-th two-level
fluctuator; and $\delta_q(t)\equiv\delta_q$ is an angular
frequency shift induced by chaotic signals. If the initial state
of the system is separable,
$\hat{\rho}(0)=\hat{\rho}_{S0}\otimes\hat{\rho}_{B0}$, we can
reduce the dynamical equation of the total system by tracing out
the degrees of freedom of the bath. The influence of the chaotic
signal $\delta_q(t)$ falls into two aspects: (1) $\delta_q(t)$
affects the angular frequency shift of the qubit induced by the
bath
\begin{eqnarray*}
\Delta\omega_q=\int_{\omega_{c_1}}^{\omega_{c_2}} d\omega
\frac{J\left(\omega\right)}{2}{\rm Im}\int_0^t
e^{i\left(\omega_q-\omega\right)\left(t-t'\right)+i\int_{t'}^t\delta_qdt^{\prime\prime}}dt';
\end{eqnarray*}
and (2), more importantly, it modifies the bath-induced
decoherence rate of the qubit which can be written as:
\begin{eqnarray*}
\Gamma_q=2\int_{\omega_{c_1}}^{\omega_{c_2}} d\omega
J\left(\omega\right){\rm Re}\int_0^t
e^{i\left(\omega_q-\omega\right)\left(t-t'\right)+i\int_{t'}^t\delta_qdt^{\prime\prime}}dt',
\end{eqnarray*}
where $J\left(\omega\right)=\sum_i
g_i^2\delta\left(\omega-\omega_i\right)$ is the spectral density
of the bath. Here, since the frequencies of the fluctuators
distribute in a finite domain, $\Delta\omega_q$ and $\Gamma_q$ are
restricted to be integrated in the finite frequency domain
$[\omega_{c1},\omega_{c2}]$. We demonstrate our results using the
zero-temperature bath.

We now come to show how the damping rate can be reduced by the frequency shift
$\delta_q(t)$, using the function $\delta_q\left(t\right)$ as a linear combination
of sinusoidal signals with small
amplitudes and high frequencies, i.e.,
$\delta_q\left(t\right)=\sum_{\alpha}
A_{d\alpha}\cos\left(\omega_{d\alpha}t+\phi_{\alpha}\right)$.
Here $\omega_{d\alpha}$ should satisfy the conditions:
$\omega_{d\alpha}\gg|\omega_{c_2}-\omega_q|$,
$|\omega_q-\omega_{c_1}|$, and $A_{d\alpha}/\omega_{d\alpha}\ll
1$. Using the Fourier-Bessel series identity~\cite{LanZhou}:
$e^{ix\sin y}=\sum_n J_n\left(x\right)e^{iny}$ with
$J_n\left(x\right)$ as the $n$-th Bessel function of the first
kind and the approximation $J_0\left(x\right)\approx1-x^2/4$ for
$x\ll 1$, we have
\begin{eqnarray*}
\int_0^t
e^{i2\omega_-\left(t-t'\right)+i\int_{t'}^t\delta_q\left(t^{\prime\prime}\right)dt^{\prime\prime}}dt'\approx
F e^{i \omega_- t} \left(\frac{\sin \omega_- t}{\omega_-}\right),
\end{eqnarray*}
where the correction factor
$F=\exp\left(-\pi\int_{\omega_{cd}}^{\infty}\frac{S_{\delta_q}\left(\omega\right)}{\omega^2}d\omega\right)$;
$S_{\delta_q}\left(\omega\right)=\sum_{\alpha}A_{d\alpha}^2\delta\left(\omega_{d\alpha}-\omega\right)/2\pi$
is the power spectrum density of the signal
$\delta_q\left(t\right)$;
$\omega_-=\left(\omega_q-\omega\right)/2$; and $\omega_{cd}$ is
the lower bound of the frequency of $\delta_q\left(t\right)$ such
that
$\omega_{cd}\gg|\omega_{c_2}-\omega_q|,\,|\omega_q-\omega_{c_1}|$.
Here, we omit the higher-order Bessel function terms because
$1/\left(i\left(\omega_q-\omega\right)+n\omega_{d\alpha}\right)\ll
1/\left(i\left(\omega_q-\omega\right)\right)$ and
$J_n\left(A_{d\alpha}/\omega_{d\alpha}\right)\ll
J_0\left(A_{d\alpha}/\omega_{d\alpha}\right)$ under the
conditions. The analysis shows that
$\delta_q\left(t\right)$ induces a correction factor $F$ for the
environment-induced frequency-shift $\Delta\omega_q$ and damping
rate $\Gamma_q$, i.e.,
\begin{eqnarray*}
\Delta\omega_q&=&F\int_{\omega_{c_1}}^{\omega_{c_2}}
d\omega\frac{J\left(\omega\right)\left[1-\cos\left(\omega_q-\omega\right)t\right]}{2\left(\omega_q-\omega\right)}=F\Delta\omega_{q0},\\
\Gamma_q\left(t\right)&=&F\int_{\omega_{c_1}}^{\omega_{c_2}}
d\omega\frac{2J\left(\omega\right)\sin\left(\omega_q-\omega\right)t}{\omega_q-\omega}=F\Gamma_{q0},
\end{eqnarray*}
where $\Delta\omega_{q0}$ and $\Gamma_{q0}$ are the
frequency-shift and damping rate when $\delta_q(t)=0$. The
correction factor $F$ may become extremely small when
$\delta_q\left(t\right)$ is a chaotic signal which has a broadband
frequency spectrum in particular in the high-frequency domain,
such that the decay rate and frequency shift can be suppressed by
a chaotic signal.

\emph{Generation of chaotic signals and suppressing $1/f$
noises.---} To show the validity of our method, as an example, we
show how to suppress the $1/f$ noises of a qubit with free
Hamiltonian $\hat{H}_q=\omega_q\hat{S}_z$ by coupling it to a
driven Duffing oscillator which is used to generate chaotic
signals, with Hamiltonian
\begin{equation}\label{Hamiltonian of the nonlinear Duffing oscillator}
\hat{H}_{\rm Duf}=\omega_o
\hat{a}^{\dagger}\hat{a}-\frac{\lambda}{4}\left(\hat{a}+\hat{a}^{\dagger}\right)^4-I\left(t\right)\frac{1}{\sqrt{2}}\left(\hat{a}+\hat{a}^{\dagger}\right)
\end{equation}
and damping rate $\gamma$, where $\hat{a}$ and $\hat{a}^{\dagger}$
are the annihilation and creation operators of the nonlinear
Duffing oscillator; $\omega_o/2\pi$ is the frequency of the
fundamental mode; $\lambda$ is the nonlinear constant; and
$I\left(t\right)=I_0\cos\left(\omega_d t\right)$ denotes the
classical driving field with amplitude $I_0$ and frequency
$\omega_d/2\pi$. We employ the interaction between the qubit and
the Duffing oscillator,
$\hat{H}_I=g_{qo}\hat{S}_z\hat{a}^{\dagger}\hat{a}$
($g_{qo}$ - coupling strength), which can be obtained, e.g., by the
Jaynes-Cummings model under the large detuning regime~\cite{Blais}.

By tracing out the degrees of freedom of the oscillator
initially in a coherent state $|\alpha\rangle$, we find that the
interaction between the qubit and the Duffing oscillator
introduces an additional factor for the non-diagonal entries of
the state of the qubit~\cite{WGWang}:
\begin{equation}\label{Non-diagonal factor induced by the dc-SQUID}
f_{01}\left(t\right)=\langle\alpha|e^{it\left(H_{\rm
Duf}+g_{qo}\hat{a}^{\dagger}a\right)}e^{-it \left(H_{\rm
Duf}-g_{qo}\hat{a}^{\dagger}a\right)}|\alpha\rangle.
\end{equation}

There are two aspects of the factor
$f_{01}\left(t\right)=e^{\Sigma_q(t)+i\Theta_q(t)}$. The phase
shift $\Theta_q(t)\approx\int_0^t\delta_q(t')dt'$, induced by
$f_{01}\left(t\right)$, is related to $\delta_q(t)$ in
Eq.~(\ref{Hamiltonian of the total system}), which can be used to
suppress the decoherence of the qubit. However, the amplitude
square
$M(t)=\left|f_{01}\left(t\right)\right|^2=e^{2\Sigma_q\left(t\right)}$,
i.e., the quantum Loschmidt echo~\cite{Peres,Jalabert}, leads to
additional decoherence effects of the qubit. Such decoherence
effects have been well studied in the literature for regular and
chaotic dynamics~\cite{WGWang,Peres,Jalabert,Vanicek,Jacquod}. Our
decoherence suppression strategy is valid when the decoherence
suppression induced by $\delta_q\left(t\right)$ is predominant in
comparison with the opposite decoherence acceleration process.

We now come to show numerical results, using
system parameters:
\begin{eqnarray}\label{Parameters}
\left(\omega_o,g_{qo},\gamma,\lambda\right)=\left(\omega_q,0.03\omega_q,0.05\omega_q,0.25\omega_q\right).
\end{eqnarray}
The bath has a $1/f$ noise spectrum (see, e.g.,
Ref.~\cite{Tsai,Shiokawa}) with $J\left(\omega\right)=A/\omega$
and $A/\omega_q=0.1$. The evolution of the coherence
$C_{xy}=\langle\hat{S}_x\rangle^2+\langle\hat{S}_y\rangle^2$ of
the qubit and the spectrum analysis of the angular frequency shift
$\delta_q\left(t\right)$ are presented in Fig.~\ref{Fig of the
decoherence suppression by chaotic device}. As shown in
Fig.~\ref{Fig of the decoherence suppression by chaotic
device}(b), (c), if we tune the amplitude $I_0$ of the sinusoidal
driving field $I\left(t\right)$ such that $I_0/\omega_q=5$ and
$30$, the signals $\delta_q\left(t\right)$ exhibit periodic and
chaotic behaviors. As shown in Fig.~\ref{Fig of the decoherence
suppression by chaotic device}(a), in the periodic regime, the
decoherence of the qubit is almost unaffected by the Duffing
oscillator. The trajectory in the periodic case (green curve with
plus signs) coincides with that of natural decoherence (black
triangle curve), as in Fig.~\ref{Fig of the decoherence
suppression by chaotic device}(a). In the chaotic regime, the
decoherence of the qubit is efficiently slowed down (see the blue
solid curve in Fig.~\ref{Fig of the decoherence suppression by
chaotic device}(a) representing the trajectory in the chaotic
case). This demonstrates that, with the increase of the
distribution of the spectral energy in the high-frequency domain,
the decoherence effects are suppressed as explained in the last
section.

\begin{figure}[h]
\centerline{
\includegraphics[width=1.72in,height=1.2in]{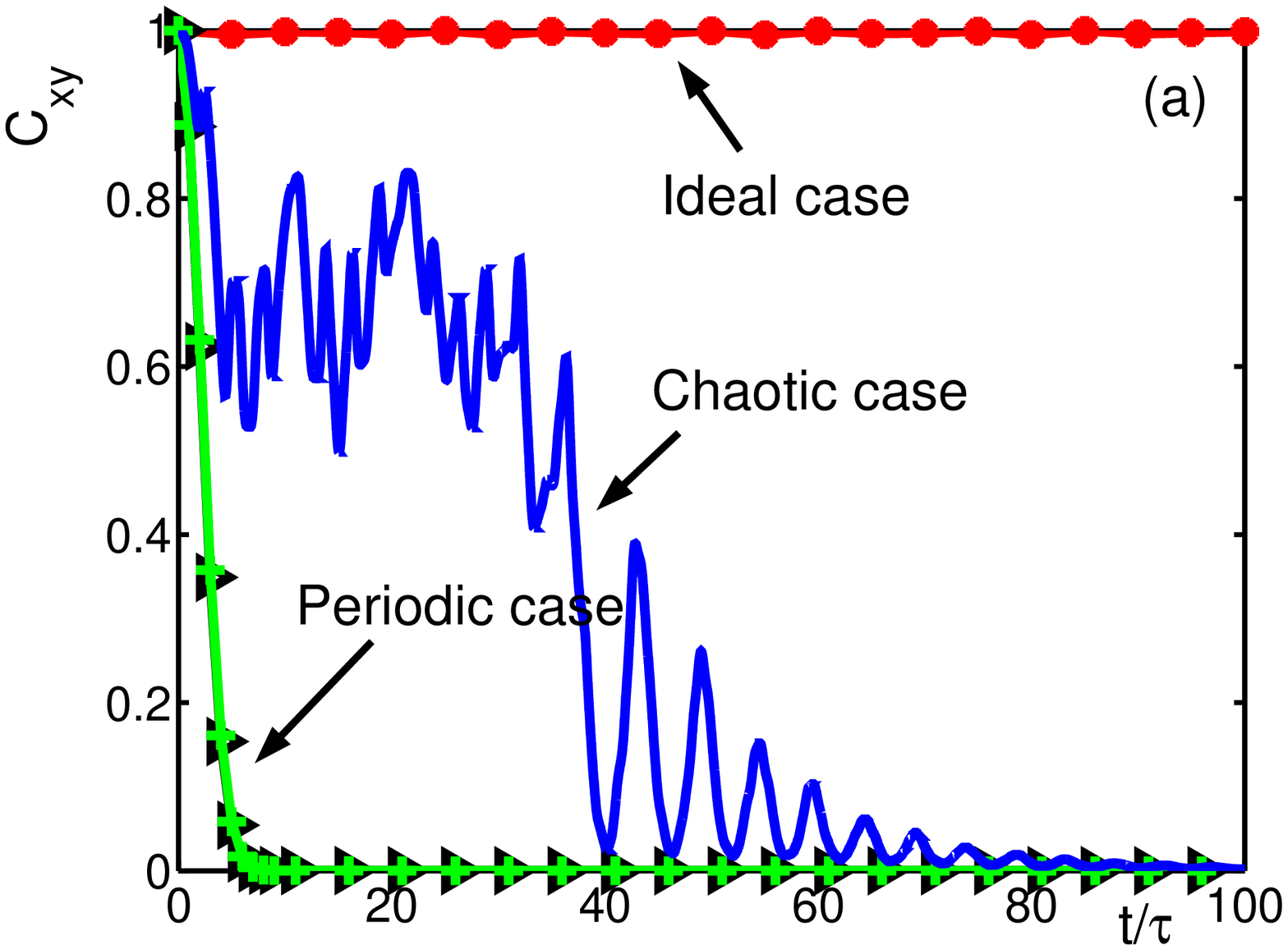}
\includegraphics[width=1.7in,height=1.2in]{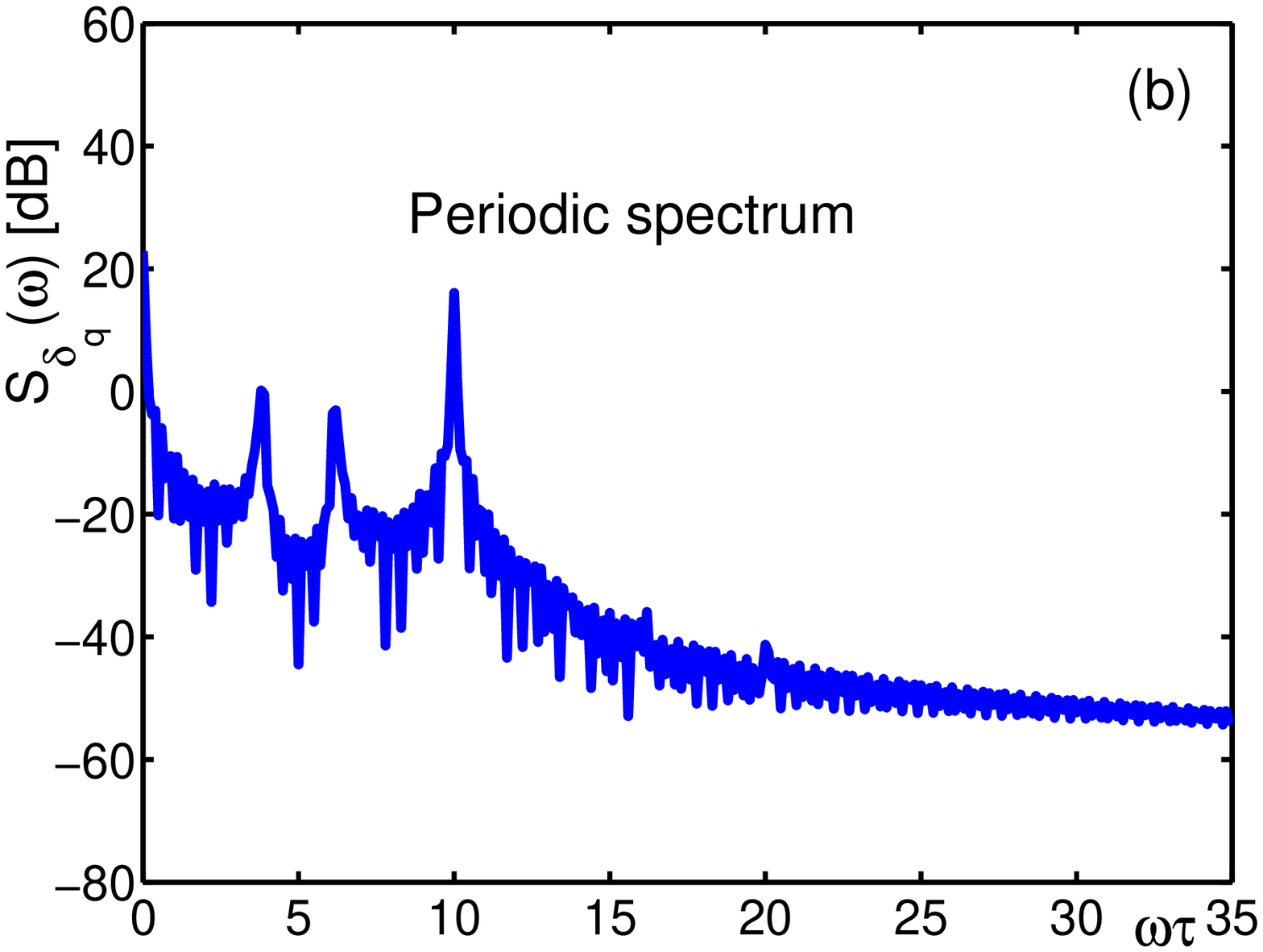}}
\centerline{
\includegraphics[width=1.7in,height=1.2in]{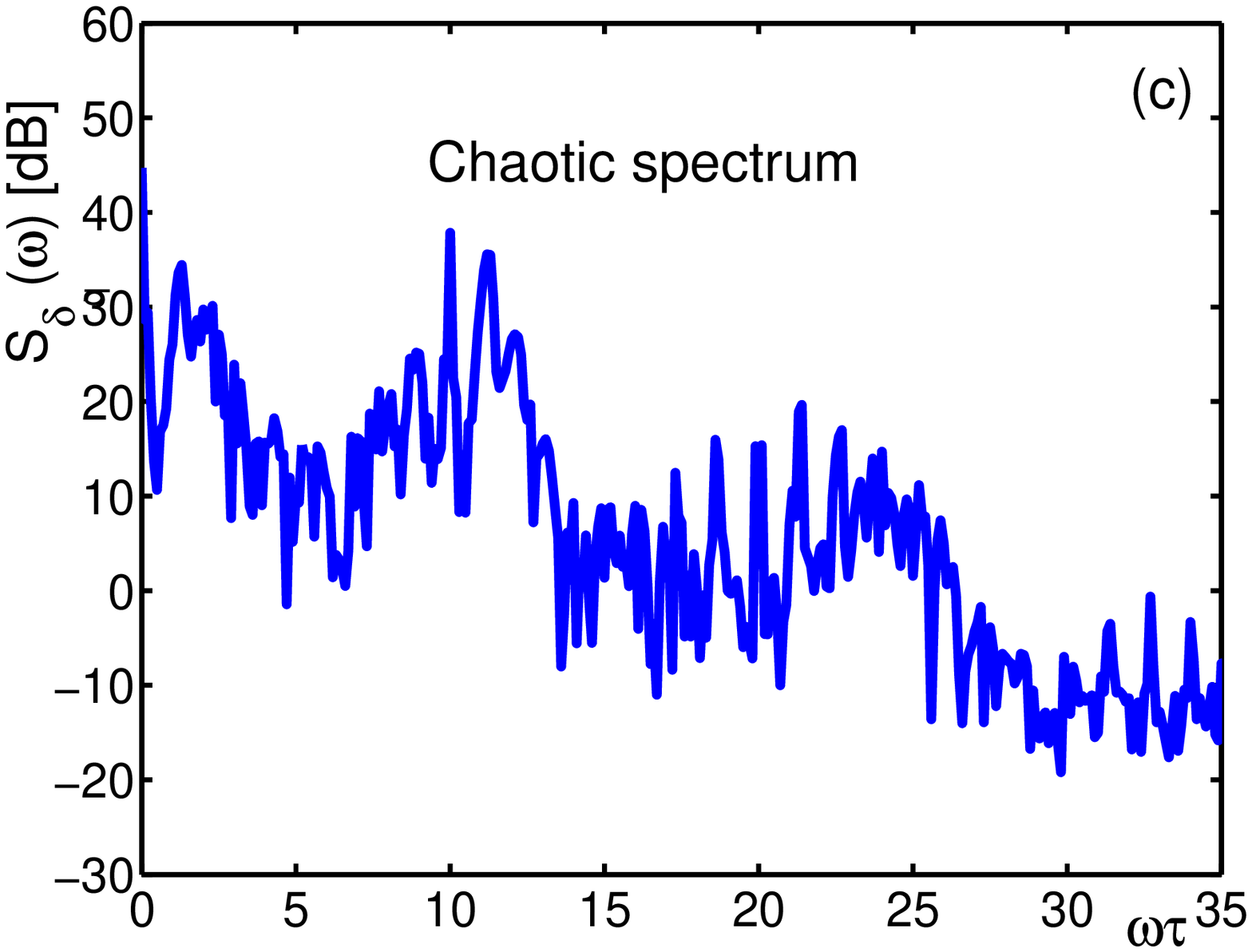}
\includegraphics[width=1.72in,height=1.2in]{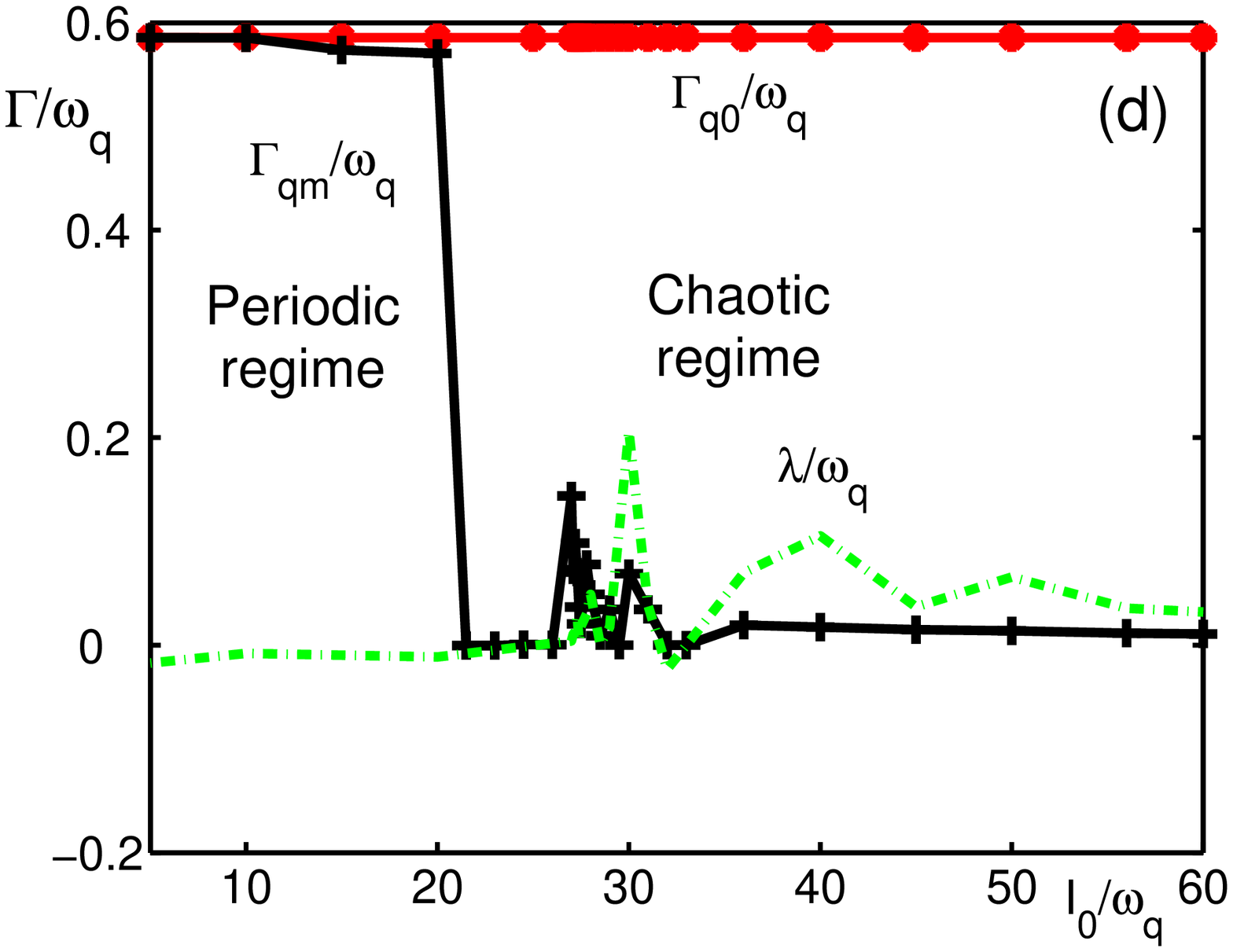}} \caption{(color online) Decoherence suppression by the auxiliary chaotic setup.
(a) the evolution of the coherence
$C_{xy}=\langle\hat{S}_x\rangle^2+\langle\hat{S}_y\rangle^2$ of
the state of the qubit, where the red asterisk curve and the black
triangle curve represent the ideal trajectory without any
decoherence and the trajectory under natural decoherence and
without corrections; and the green curve with plus signs and the
blue solid curve denote the trajectories with $I_0/\omega_q=5$ and
$30$. With these parameters, the dynamics of the Duffing
oscillator exhibits periodic and chaotic behaviors.
$\tau=2\pi/\omega_q$ is a normalized time scale. (b) and (c) are
the energy spectra of $\delta_q(t)$ with $I_0/\omega_q=5$ (the
periodic case) and $30$ (the chaotic case). The energy spectrum
$S_{\delta_q}\left(\omega\right)$ is in unit of decibel (dB). (d)
the normalized decoherence rates $\Gamma/\omega_q$ versus the
normalized driving strength $I_0/\omega_q$.}\label{Fig of the
decoherence suppression by chaotic device}
\end{figure}

We compare in Fig.~\ref{Fig of the decoherence suppression by
chaotic device}(d) the average natural decoherence rate
$\Gamma_{q0}$ and modified decoherence rate $\Gamma_{qm}$ of the
qubit versus different strengths $I_0$ of the driving field.
Figure~\ref{Fig of the decoherence suppression by chaotic
device}(d) shows that, when the strength $I_0$ of the driving
field increases, the decoherence process is efficiently slowed
down. It is interesting to note that there seems to exist a phase
transition around $I_0/\omega_q=20$, i.e., a sudden change of the
modified decoherence rate $\Gamma_{qm}$ (see the black solid curve
with plus signs). It is noticable that, around this point, the
dynamics of the Duffing oscillator enters the chaotic regime which
is indicated by a positive Lyapunov exponent (see the green
dash-dotted curve in Fig.~\ref{Fig of the decoherence suppression
by chaotic device}(d)). The modified decoherence rate
$\Gamma_{qm}$ changes dramatically in the parameter regime
$I_0/\omega_q\in\left[20,35\right]$ which is the soft-chaos regime
of the Duffing oscillator. When the dynamics of the Duffing
oscillator enters the hard-chaos regime at $I_0/\omega_q\approx
35$, the modified decoherence rate $\Gamma_{qm}$ is stabilized at
a value much smaller than the natural decoherence rate
$\Gamma_{q0}$. The simulation results show that the decoherence of
the qubit is efficiently suppressed by our proposal, even if there
exists an additional decoherence introduced by the auxiliary
chaotic setup. Further calculations show that the modified
decoherence rate $\Gamma_{qm}$ in the chaotic regime is roughly
$100$ times smaller than the unmodified decoherence rate
$\Gamma_{q0}$, meaning that the decoherence time of the qubit can
be prolonged $100$ times.

\emph{Experimental feasibility in superconducting circuits.---}
Our general study can be demonstrated using the solid state
quantum devices, e.g., the superconducting qubit system, as sketched
in Fig.~\ref{Fig of the protecting superconducting circuit}. It is
similar to the widely used qubit readout circuit~\cite{Siddiqi}, but works
in a quite different parameter regime. In this
superconducting circuit, a single Cooper pair box (SCB) is coupled
to a dc-SQUID consisting of two Josephson junctions with
capacitances $\tilde{C}_J$ and Josephson energies $\tilde{E}_J$
and a paralleled current source. The SCB is composed of two
Josephson junctions with capacitances $C_J$ and Josephson energies
$E_J$. The difference between the circuit in Fig.~\ref{Fig of the
protecting superconducting circuit} and the readout circuit in
Ref.~\cite{Siddiqi} is that the rf-biased Josephson junction is
replaced by a dc-SQUID - the chaotic setup.

\begin{figure}[h]
\includegraphics[bb=75 223 592 552, width=5.5 cm,
clip]{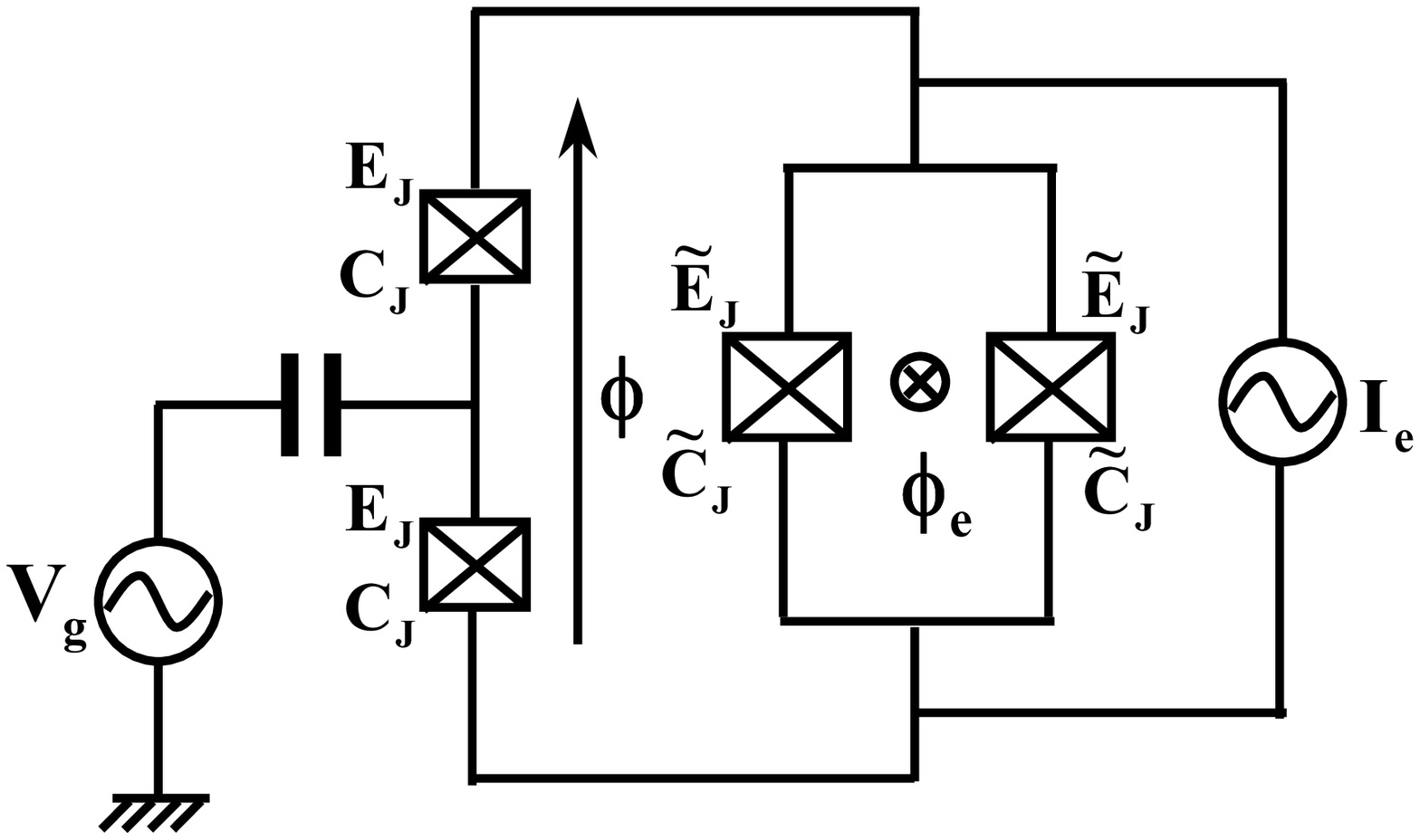} \caption{(color online) Schematic diagram of the
decoherence suppression superconducting circuit in which a SCB is
coupled to a current-biased dc-SQUID.}\label{Fig of the protecting
superconducting circuit}
\end{figure}

The Hamiltonian of the circuit shown in Fig.~\ref{Fig of the
protecting superconducting circuit} can be written as:
\begin{eqnarray}\label{Hamiltonian of the protecting superconducting circuit}
\hat{H}&=&E_C\left(\hat{n}-n_g\right)^2-2E_J\cos\frac{\hat{\phi}}{2}\cos\hat{\theta}+\tilde{E}_C
\hat{\tilde{n}}^2\nonumber\\
&&-2\tilde{E}_J\cos\frac{\phi_e}{2}\cos\hat{\phi}-\phi_0 I_e
\hat{\phi},
\end{eqnarray}
where $E_C=2e^2/\left(C_g+2C_J\right)$ is the charging energy of
SCB with $C_g$ as the gate capacitance; $n_g=-C_gV_g/2e$ is the
reduced charge number, in unit of the Cooper pairs, with $V_g$ as
the gate voltage; $\hat{n}$ is the number of Cooper pairs on the
island electrode of SCB with $\hat{\theta}$ as its conjugate
operator; $\tilde{E}_C=e^2/\tilde{C}_J$ is the charging energy of
the dc-SQUID; $\hat{\tilde{n}}$ is the charge operator of the
dc-SQUID and $\hat{\phi}$ the conjugate operator; and $\phi_e$ and
$I_e$ are the external flux threading the loop of the dc-SQUID and
the external bias current of the dc-SQUID. Here, we consider a
zero external flux threading the loop of the coupled SCB-dc SQUID
system. In this case, the phase drop across the SCB is equal to
the phase drop across the dc-SQUID $\tilde{\phi}$. We further
introduce the ac gate voltage $V_g=V_{g0}\cos\left(\omega_g
t\right)$ with amplitude $V_{g0}$ and angular frequency
$\omega_g$. With the condition that $C_gV_{g0}E_C/2e\ll
\omega_q=E_J-\omega_g$, the SCB works near the optimal point, and
we only need to worry about the relaxation of the SCB. By
expanding the Hamiltonian of the SCB in the Hilbert space of its
two lowest states and leaving the lowest nonlinear terms of
$\hat{\phi}$ in the rotating frame, we can obtain the effective
Hamiltonian $\hat{H}_{\rm eff}=\hat{H}_q+\hat{H}_{\rm
Duf}+\hat{H}_I$ discussed in the foregoing section. This dc-SQUID,
acting as the auxiliary Duffing oscillator, can be used to
suppress low frequency $1/f$ noises of the qubit. Using the
experimentally accessible parameters as shown in the caption of
table~\ref{Table of decoherence suppression under various noises},
we show the decoherence suppression effects for low frequency
$1/f$, high frequency Ohmic ($J\left(\omega\right)=\omega
e^{-\omega/5\omega_q}$), sub-Ohmic
($J\left(\omega\right)=\omega^{1/2} e^{-\omega/5\omega_q}$), and
super-Ohmic ($J\left(\omega\right)=\omega^2
e^{-\omega/5\omega_q}$) noises. All simulations are summarized in
table~\ref{Table of decoherence suppression under various noises}.
It is found that our method works equally well for different types
of noises. The numerical simulations manifest that our strategy is
independent of the sources and frequency domains of the noises.
The final modified decoherence rates for these different noises
are almost the same because the decoherence effects induced by the
environmental noises are all greatly suppressed, and thus the
modified decoherence rates of the qubit are mainly caused by the
auxiliary chaotic setup, i.e., the dc-SQUID. It is also shown in
table~\ref{Table of decoherence suppression under various noises}
that the modified decoherence rate $\Gamma_{qm}/2\pi$ of the qubit
can be reduced to $5$ kHz. This low decoherence rate corresponds
to a long decoherence time $T_1=T_2\approx 200\,\mu$s. The
magnitude is one-order longer than the decoherence time of the
superconducting qubits realized in experiments (see, e.g.,
Ref.~\cite{Tsai}).


\begin{table}
\begin{tabular}{|c|c|c|c|c|}
  \hline
  Type of noises & Frequency domain & $\bar{\Gamma}_{q0}$ & $\bar{\Gamma}_{qm}$ & $T_1=T_2$ \\
  \hline
  1/f noise & [$10$ kHz, $1$ MHz] & $0.58$ MHz & $5.4$ kHz & $187\,\mu$s \\
  Ohmic & $[2\omega_q/3,3\omega_q/2]$ & $0.35$ MHz & $5.4$ kHz & $187\,\mu$s \\
  Sub-Ohmic & $[2\omega_q/3,3\omega_q/2]$ & $0.35$ MHz & $5.4$ kHz & $187\,\mu$s \\
  Super-Ohmic & $[2\omega_q/3,3\omega_q/2]$ & $0.36$ MHz & $5.4$ kHz & $187\,\mu$ s \\
  \hline
\end{tabular}\caption{Decoherence suppression against various
noises for experimentally accessible parameters: $E_J/2\pi=5$ GHz,
$\omega_g/2\pi=4.999$ GHz, $\tilde{E}_C/2\pi=0.188$ MHz, and
$\tilde{E}_J\cos\frac{\phi_e}{2}/2\pi=12.032$ MHz.}\label{Table of
decoherence suppression under various noises}
\end{table}

\emph{Conclusion.---} In conclusion, we propose a strategy to
prolong the decoherence time of a qubit by coupling it to a
chaotic setup. The broad power distribution of the auxiliary
chaotic setup in particular in the high-frequency domain helps us
to suppress various non-Markovian noises, e.g., low-frequency
$1/f$ noise, high-frequency Ohmic, sub-Ohmic, and super-Ohmic
noises, and thus freeze the state of the qubit even if we consider
the additional decoherence induced by the chaotic setup. We find
that the decoherence time of the qubit can be efficiently
prolonged approximately $100$ times in magnitude. We believe that
our strategy is feasible, in particular for a coupled SCB-SQUID
system, and also gives a new perspective for the reversibility and
irreversibility induced by nonlinearity.

J. Zhang would like to thank Dr. H.~T. Tan and Dr. M.~H. Wu for
helpful discussions. We acknowledge financial support from the
National Natural Science Foundation of China under Grant Nos.
60704017, 10975080, 61025022, 60904034, 60836001, 60635040. T.~J.
Tarn would also like to acknowledge partial support from the U. S.
Army Research Office under Grant W911NF-04-1-0386. L. A. Wu has
been supported by the Ikerbasque Foundation Start-up, the Basque
Government (grant IT472-10) and the Spanish MEC (Project No.
FIS2009-12773-C02-02).
\\[0.2cm]

\end{document}